\documentstyle[12pt,aaspp4]{article} 

\def\gtrsim{\mathrel{\hbox{\rlap{\hbox{\lower4pt\hbox{$\sim$}}}\hbox{$>$}}}}
\def\lesssim{\mathrel{\hbox{\rlap{\hbox{\lower4pt\hbox{$\sim$}}}\hbox{$<$}}}}
\begin{document}
  
\title{Thermal Water Vapor Emission from Shocked Regions in Orion$^*$}
  
\author{Martin Harwit$^1$, David A. Neufeld$^2$, Gary J. Melnick$^3$, and Michael J.
Kaufman$^4$}
  
\vskip 0.5 true in {\parskip 6pt \noindent{$^1$ 511 H Street S.W., Washington, DC
20024--2725}
  
\noindent{$^2$ Department of Physics \& Astronomy,  The Johns Hopkins University, 3400
North Charles Street,  Baltimore, MD 21218}
  
\noindent{$^3$ Harvard-Smithsonian Center for Astrophysics, 60 Garden Street, Cambridge,
MA 02138}
  
\noindent{$^4$ NASA Ames Research Center, MS 245-3,  Moffett Field, CA 94035--1000}
\vskip 0.5 true in \noindent{$^*$ Based on observations with ISO, an ESA project with
instruments funded by ESA Member States (especially the PI countries: France, Germany, the
Netherlands and the United Kingdom) with the participation of ISAS and NASA.} \vskip 0.5
true in}
  
\keywords{infrared: ISM: lines and bands --- ISM: abundances --- ISM: individual (Orion) ---
ISM: molecules --- molecular processes }
  
\begin{abstract}
  
Using the Long Wavelength Spectrometer (LWS) onboard the Infrared Space Observatory
(ISO), we have observed thermal water vapor emission from a roughly circular field of view 
approximately 75 arc seconds in diameter centered on the Orion BN-KL region. The
Fabry-Perot line strengths, line widths, and spectral line shifts observed in eight transitions
between 71 and 125$\mu$m show good agreement with models of thermal emission arising
from a molecular cloud subjected to a magnetohydrodynamic C-type shock. Both the breadth
and the relative strengths of the observed lines argue for emission from a shock rather than
from warm quiescent gas in the Orion core. Though one of the eight transitions appears
anomalously strong, and may be subject to the effects of radiative pumping, the other seven
indicate an H$_2$O/H$_2$ abundance ratio of order $5\times 10^{-4}$, and a corresponding
gas-phase oxygen-to-hydrogen abundance ratio of order $4\times 10^{-4}$. Given current
estimates of the interstellar, gas-phase, oxygen and carbon abundances in the solar vicinity,
this value is consistent with theoretical shock models that predict the conversion into water of
all the gas-phase oxygen that is not bound as CO. The overall cooling provided by rotational
transitions of H$_2$O in this region appears to be comparable to the cooling through
rotational lines of CO, but is an order of magnitude lower than cooling through H$_2$
emission. However, the model that best fits our observations shows cooling by H$_2$O and
CO dominant in that portion of the post-shock region where temperatures are below $\sim
800$\,K and neither vibrational nor rotational radiative cooling by H$_2$ is appreciable.
  
\end{abstract}

\section {Introduction}
  
The molecular hydrogen shock in the Orion BN-KL region has been intensively studied since
the seminal work of Kwan and Scoville (1976) revealed emission from carbon monoxide with
line widths of order 100 km s$^{-1}$. Following this, Beckwith {\it et al.} (1978) identified
the rovibrational transitions of molecular hydrogen emanating from this same region.
Somewhat later, Beck (1984) provided a map of the region in the (v = 0--0 S(2)) 12.3$\,\mu$m
line, while  studies by Watson {\it et al.} (1980), Stacey {\it et al.} (1983), and Watson {\it et
al.} (1985) identified highly excited far-infrared rotational transitions from CO involving
rotational levels ranging from $J = 15$ up to $J = 34$. Watson (1982), Viscuso {\it et al.}
(1985), and Melnick {\it et al.} (1990) also studied OH transitions, with a view to
understanding the dynamics and chemistry of the shocked and post-shock domains. 
  
More recently, Sugai {\it et al.} (1994) have provided a detailed map of the molecular
hydrogen emission. They mapped a $4'\times 4'$ region with a spatial resolving power of 5.3
arc seconds in the H$_2$ (v = 1--0 S(1)) transition and found substantial emission from an
elongated patch roughly two minutes of arc in length and 1.5 arc minutes wide. In this region
they resolved nine separate emission peaks in a structure whose outlines are roughly bipolar,
straddling the central source IRc2. While many smaller peaks abound, the overall structure is
generally interpreted in terms of a bipolar wind emanating from IRc2 and plowing into an
ambient H$_2$ cloud.
  
Most theoretical studies of the Orion shocked region (Draine \& Roberge, 1982; Chernoff,
Hollenbach \& McKee, 1982; Neufeld \& Melnick, 1987) have concluded that a rich emission
spectrum from thermally excited water vapor must be playing a significant role in cooling the
gas. However, with few exceptions --- such as observations of maser emission from extremely
dense regions or observations of relatively low-excitation lines of the isotopomers HDO and
H$_2\,^{18}$O (e.g. Zmuidzinas {\it et al.} 1995) --- the direct detection of water vapor
emission has not been possible. Though thermal water vapor emission, largely confined to the
far-infrared, has long been sought, telluric water vapor is an efficient absorber in the mid- and
far-infrared, and blocks precisely those wavelengths at which interstellar shocks are expected
to emit the bulk of their radiation. While Cernicharo {\it et al.} (1994) have reported detection
of the 183.3 GHz radio frequency 3(1,3) -- 2(2,0) transition, decisive observations had to
await the launch of mid- and far-infrared spectral instrumentation into space. The Infrared
Space Observatory, ISO, placed into earth orbit in November 1995, has provided a first
opportunity to study these shocks systematically. Here, we provide evidence that this emission
has now been reliably observed. 
  
\section{Observations}
  
On October 6, 1997, we observed the Orion BN-KL region with ISO (cf Kessler {\it et al.},
1996) from 06:02:47 to 07:58:05 UT for a total of 6918 seconds. All observations were made
in the Long Wavelength Spectrometer's Fabry-Perot (LWS/FP) mode (Clegg {\it et al,} 1996).
The instrument's roughly circular field of view was centered on the epoch 2000 coordinates
$5h\;35m\;14.2s,\ -5^{\circ}\; 22'\; 23.3"$.
  
We obtained data on eight H$_2$O lines. These detections required the use of four different
detectors, each having its own roughly elliptical beam size --- $70\times 68$ arc seconds at
125 and 121 $\mu$m, $77\times 71$ at 99.5, 95.6 and 90 $\mu$m, $82\times 76$ at 83 and 82
$\mu$m, and $83\times 79$ at 72 $\mu$m. The fields of view, therefore, differed by $\pm
19\%$.
  
In the wavelength range from 72 to 125 $\mu$m the LWS/FP's spectral resolving power
gradually increases from $\sim 7,000$ to a peak of $\sim 9,800$ at 95 $\mu$m, before slowly
dropping to $\sim 9,500$ at 125$\mu$m. We repeatedly stepped the Fabry-Perot over a
wavelength range that spanned 5 resolution elements on either side of each line. This made for
a total of 11 resolution elements, each of which was sampled at 8 equally spaced positions
within the element, in what is designated as the ``rapid scanning" mode of operation, each
measurement lasting only 0.5 sec. Strong signals were observed in all eight lines (see Figure
1). 
  
\section{Results}
  
The main observational results are listed in Table 1. The first two columns, respectively,
identify the transitions observed and indicate whether the line is emitted by ortho- or
para-H$_2$O. The third column gives the wavelength. The fourth column lists the line center
displacement in terms of velocity with respect to the local standard of rest. Like all the other
observational parameters listed in Table 1, this was determined from a Gaussian line fit. Since
the spectral resolving power of the Fabry-Perot spectrometer ranges from 9,000 to 7,000, the
lines necessarily are broader than 33 to 43 km s$^{-1}$. The relatively small spread of central
line velocities observed, $\pm 8$ km s$^{-1}$, probably is due to noise, and suggest that all
of the lines are emitted from one and the same region. Inspection of Figure 1, where the
observed line profiles are displayed, visually confirms the quality of the data and shows the
small statistical errors that arise in comparing the many scans taken over each of the lines.
Column 5 of Table 1 gives the line widths. These widths are of the order of 60 km s$^{-1}$
full width at half maximum, significantly exceeding the minimum line width of 33 to 43 km
s$^{-1}$ determined by the spectral resolving power. They are consistent with an intrinsic
line-of-sight velocity distribution that is Gaussian with full-width-half-maximum $\sim 35$ km
s$^{-1}$, convolved with the Fabry-Perot's Lorentzian profile at the above-stated resolving
power. This estimate of the intrinsic line width is supported by additional observations toward
Orion Peak 1 to be reported elsewhere. These made use of ISO's Short Wavelength
Spectrometer, whose resolving power of $\sim 31,000$ displayed a 42 km s$^{-1}$
full-width-half-maximum velocity distribution in the $4_{32}- 3_{03}$ H$_2$O emission
line at 40.69 $\mu$m.
  
Column 6 of Table 1 shows the total flux measured in each line. Columns 7 and 8 involve
theoretical modeling and are discussed below. Column 9, the last column, provides the
detected continuum level. It is substantial, and underscores the importance of correct
continuum subtraction. At 100\,$\mu$m a flux of $\sim 1.5\times 10^5$ Jy corresponds to
$\sim 5\times 10^{-17}$ W cm$^{-2}$ per resolution element, comparable to the total line
flux.
  
The calibration of the LWS/FP is a complex procedure that continues to be refined. It consists,
in first place, of a calibration of the LWS detector response in the instrument's grating mode.
This is periodically monitored through the use of five infrared illuminators incorporated into
the instrument, and by observing selected astronomical calibration sources. The response in
the Fabry-Perot mode, for which the grating serves as an order sorter, is then deduced
primarily from a preflight calibration of the Fabry-Perot etalon's added transmission losses.
  
As seen in the line plots of Fig. 1, the continuum appears to be steep for some of the lines. A
variety of factors contribute to these slopes. They include changes in the transmission function
of the blocking filters and grating, a changing detector response function, and interference
effects within the instrument that lead to sizeable fringe amplitudes. To the extent that our
continuum levels agree with continuum observations obtained by others, they nevertheless
provide us with a means of ascertaining the reliability of the line fluxes we list. 
  
Werner {\it et al.} (1976) mapped the same region of Orion we observed, with a field of view
one arc minute in diameter, and described its spectrum as a blackbody at 70 K with a $20\,\mu
{\rm m}/\lambda$ emissivity. We find rough agreement in the spectral distribution of our
continuum values with the spectral shape they describe. Moreover, the peak flux Werner {\it
et al.} observed at 100\,$\mu$m with a 40\,$\mu$m bandwidth was $9\times 10^4$\,Jy. The
continuum flux levels listed in Table 1 were all obtained with larger fields of view, but if we
roughly prorate our observed flux to the smaller field of view of Werner {\it et al.}, taking into
account that their maps show a rapid decrease in flux away from the peak, we derive
respective fluxes of $ \sim 0.64,\ \sim 1.25$ and $\sim 1.57\times 10^5$\,Jy at 121.7, 99.49
and 82.03\,$\mu$m. Since these three values provide a rough sampling of their wavelength
range, we can average them to obtain a mean value $\sim 1.15\times 10^5$\,Jy or 25\%
higher. Werner {\it et al.} estimated the uncertainties in their absolute flux levels at $\pm
20\%$. Summing the two uncertainties of 25 and 20\% in quadrature, we obtain an uncertainty
in the continuum fluxes listed in Table 1, of $\sim \pm35\%$. On the basis of this comparison
we estimate our line fluxes to be similarly uncertain by $\sim \pm 35\%$. As the Long
Wavelength Spectrometer characteristics become better established in the next few months,
this error budget may be revised.
  
\section{The Model}
  
The luminosity of water vapor emission from the Orion shock was predicted by Kaufman \&
Neufeld (1996) [hereafter KN] and in Kaufman's PhD thesis (1995). Their model solved for
the equilibrium populations of the lowest 179 rotational levels of ortho-H$_2$O and for the
lowest 170 rotational levels of para-H$_2$O. The model included cooling due to rovibrational
transitions of H$_2$O, H$_2$ and CO, and due to dissociation of H$_2$. Cooling through
gas-grain collisions was also included. 
  
Water vapor in shocks is formed through reactions of atomic oxygen with molecular hydrogen
to form OH radicals, which subsequently react with hydrogen molecules to form water vapor
(Elitzur \& de Jong, 1973; Elitzur \& Watson, 1978). Once the shock-generated temperatures
exceed $\sim 400$ K, the KN model predicts that all the oxygen not already incorporated in
CO will be converted into H$_2$O. This is shown by computations that predict the relative
abundances of the primary oxygen-bearing species, H$_2$O, OH, O, and O$_2$ throughout
the shock. Expected spectral line emission from each of these species is computed using an
escape probability formalism. The calculations included emission from the lowest 60
rotational states of $^{12}$CO as well as $^{13}$CO, and the lowest 21 rotational states of
H$_2$. For both molecular hydrogen and water vapor, the ortho/para ratio was assumed to be
3:1. Vibrational transitions from states up to v = 2 for CO and H$_2$ were included, as were
$\nu_2$ band transitions of H$_2$O.
  
In order for a shock not to dissociate interstellar molecules, the cloud through which it passes
must be able to cool itself appreciably more quickly than it is heated by the shock. This will
generally happen only if a magnetic field is present and pre-compresses the medium in what
Draine (1980) has termed a continuous or ``C-type" shock. This more gradual compression
becomes possible because the Alfv\'{e}n speed exceeds the speed of sound in the cloud. A
magnetic precursor penetrates the cloud before the arrival of a J-type shock front in the wake
of which temperatures and densities could sharply jump to dissociate the molecules.
  
KN considered a C-shock and predicted the H$_2$O emission from Orion assuming the
gas-phase oxygen and carbon abundances of Pollack {\it et al.} (1994), which were premised
on Solar System elemental abundances, and a model of the shocked region in Orion based on
data primarily obtained in CO and H$_2$ observations. These indicated a best fit for a shock
velocity $v_s = 37\,$km\,s$^{-1}$ and a preshock H$_2$ density of $10^5\, {\rm cm}^{-3}$.
These values were constrained by the observed line-strength ratios of two pairs of spectral
lines --- the CO (J=34--33) and (J=21--20) lines and the H$_2$ (v = 1--0 S(1)) and (v = 2--1
S(1)) lines.
  
The shock velocity and hydrogen density do not depend on the assumed oxygen abundance,
but the expected water vapor density does. Based on the work of Pollack {\it et al.} (1994), KN
assumed the gas phase abundances of oxygen and carbon nuclei to be, respectively,
$5.45\times 10^{-4}$ and $1.2\times 10^{-4}$ relative to hydrogen nuclei. This led to the
prediction that the water vapor abundance would be as high as n(H$_2$O)/ n(H$_2$) =
$8.5\times 10^{-4}$, and that the H$_2$, H$_2$O, and CO lines, respectively should radiate
away 75, 21 and 4\% of the shock energy. More recent data by Cardelli {\it et al.} (1996)
[hereafter CMJS], however, indicate that the gas phase in diffuse interstellar clouds within 600
pc of the Sun exhibits a remarkably constant abundance of carbon and oxygen, with n(O)/n(H)
= $3.16\times 10^{-4}$ and n(C)/n(H) = $1.4\times 10^{-4}$. In a preliminary calculation
applied to Orion, we find that  this substantially lower oxygen abundance predicts a
correspondingly reduced water vapor concentration, n(H$_2$O)/ n(H$_2$) $\sim 3.5\times
10^{-4}$, and a drop in water vapor cooling that is close to proportional to the drop in
abundance. The H$_2$ : H$_2$O : CO cooling ratios thus become 88 : 8 : 4.
  
To determine the expected H$_2$O flux from Orion, we still need to know the value of a
projection parameter which KN call $\Phi$. It is the ratio of the actual surface area of the
shock(s) to the projected area of the beam at the distance of the source. For a complex region
the field of view may contain $n$ shocks and the function $\Phi$ is summed over all of them.
For a beam-filling, planar shock $\Phi = 1$; for a beam-filling, spherical shock its value is
$\Phi = 4$. 
  
Observationally, the value of $\Phi$ is derived as the ratio of the radiant energy observed, to
the mechanical inflow energy expected for a beam-filling planar shock seen face-on. For the
Orion shock, we infer the value of $\Phi$ from an estimate of the total surface brightness in
our 75 arc second field of view, summed over all H$_2$ emission lines. To this end, we use
Beck's (1984) H$_2$(v = 0--0 S(2)) intensity averaged over the observed field of view as a
tracer for the total H$_2$ emission --- much of which is not directly observable because it is
extinguished by ambient dust. Beck's data averaged over our field of view give a 12.3\,$\mu$m
intensity of $\sim 1.6\times 10^{-3}\,{\rm erg\, cm}^{-2}{\rm \, s}^{-1}\,{\rm sr}^{-1}$,
when corrected by a factor of $\sim 2$ for 0.75 mag of extinction. The KN model indicates
that the 12.3\,$\mu$m flux needs to be multiplied by a factor of $\sim 530$, under the assumed
density and shock velocity conditions, to yield a total intensity summed over all H$_2$
spectral lines. Though this multiplier is large, we have confidence that it is quite accurate and
that it reliably implies a total H$_2$ intensity of $\sim 0.85 \,{\rm erg\, cm}^{-2}{\rm \,
s}^{-1}\,{\rm sr}^{-1}$. Assuming that this is 88\% of the total cooling, we would expect a
total radiated flux from all species of $\sim 0.97 \,{\rm erg\, cm}^{-2}{\rm \, s}^{-1}\,{\rm
sr}^{-1}$. This has to be compared to the kinetic energy inflow, which is nmv$^3/8\pi \sim
0.92 \,{\rm erg\,cm}^{-2}{\rm \, s}^{-1}\,{\rm sr}^{-1}$. The ratio of these two quantities
yields $\Phi \sim 1.05$. 
  
We are now in a position to compare the observed H$_2$O fluxes to the predicted. Table 11
of KN predicts the flux in a field of view 44 arc seconds in diameter for a shock with $\Phi\sim
3$. This field of view subtends a solid angle 0.34 times that of our $\sim 75$ arc second field.
The H$_2$O abundances predicted using the gas-phase abundances from CMJS are only 0.41
times as high as the values KN had assumed. These three effects partially cancel, but we have
to divide the values listed in Table 11of KN by 2.5 in order to apply their results to our
observations with a 75 arc second beam.
  
Column 7 of Table 1 lists the ratios of the observed fluxes to the adjusted KN predictions
based on the CMJS abundances. Given the wide range of energies $E$ at which the upper
levels for the transitions lie, $300\lesssim E \lesssim 800$ K, it is unlikely that the
observed-to-predicted line ratios would accidentally happen to fall into the narrow range in
column 7 for all but one of the eight lines observed. Our abundance estimate assumes that the
water line emission originates entirely in the shocked gas component, and not in a lower
temperature region. This is confirmed both by the line-strength ratios and by the observed line
widths. Rigorous calculations, which we do not present here, confirm that the predicted
H$_2$O fluxes scale almost linearly with H$_2$O abundance, though the average value of the
actual-to-predicted line flux in column 7 drops by about 6\% and the inferred abundances
correspondingly rise by $\sim 6\%$. For the substantial uncertainties in our flux calibrations,
in our estimate of $\Phi$, and in the abundance estimates, this agreement is reasonable. The
deviant 121.7\,$\mu$m line flux may reflect the neglect of radiative pumping by dust
continuum radiation in the KN model. Preliminary calculations show that pumping by the
ambient radiation field of Werner {\it et al.} raises the flux in this one line and leaves the other
transitions essentially unaffected. Disregarding the anomalous 121\,$\mu$m flux, we see that
the observed values are roughly 30\% higher than the predicted values.
  
\section{Discussion}
  
Molecular shocks are important not only for their intrinsic interest, but also because current
views assume that star formation may well be triggered by shock compression followed by
rapid cooling. The cooling needs to be rapid in order to prevent a quasi-elastic bounce which
would permit shock-compressed regions to rebound to their original dimensions. If a shocked
region is able to radiate away a substantial fraction of its energy during the traversal time of
the shock, it will remain compressed. Even if it is not sufficiently dense at this stage to enter
protostellar collapse, it will be poised to contract further if subjected to subsequent shocks. 
  
We can estimate the integrated water vapor line flux from Orion summed over all transitions
and compare it to the power radiated away by H$_2$ and CO. To estimate the total water
vapor emission we can sum the flux both from lines that we have observed and lines whose
strengths we infer by applying the KN model. This leads us to deduce a total water vapor
emission of $\sim 0.11\,$erg cm$^{-2}$ s$^{-1}$ sr$^{-1}$ normalized to a 75 arc second
diameter solid angle. The beam-averaged CO flux from the region is $\sim 0.13$ erg
cm$^{-2}$ s$^{-1}$ sr$^{-1}$ obtained from the observational data cited in Stacey {\it et
al.}( 1983), normalized to an assumed beam size of 60 arc seconds --- slightly smaller than the
beam size used for our H$_2$O observations. This value could, however, drop by $\sim 35\%$
if the shocked region subtended a substantially smaller solid angle than our 75 arc second
beam. The H$_2$O and CO cooling of the shocked region are therefore comparable; but both
are an order of magnitude lower than the  total H$_2$ emission. 
  
\section{Conclusions}
  
We have observed water vapor emission from the shocked region in Orion, and find that the
detected flux is consistent with a model proposed by Kaufman \& Neufeld (1996), which
inferred a shock velocity of $\sim 37$ km s$^{-1}$ and a pre-shock H$_2$ density of $\sim
10^5$ cm$^{-3}$. However the model can only be correct if we assume a water vapor
abundance n(H$_2$O)/n(H$_2$)$\sim 5\times 10^{-4}$ and a corresponding interstellar
gas-phase oxygen abundance n(O)/n(H)$\sim 4\times 10^{-4}$, in agreement with values cited
by CMJS, in place of a substantially higher Solar System abundance inferred from Pollack {\it
et al} (1994). Water vapor cooling in the shock is comparable to CO cooling, but amounts to
only $\sim 10\%$ of the total cooling provided by molecular hydrogen emission.
Nevertheless, the KN model shows that H$_2$O and CO should dominate in that portion of
the post-shock region where temperatures are $\lesssim 800$\,K and neither vibrational nor
rotational radiative cooling of H$_2$ is appreciable. 
  
We wish to thank Steven Lord at IPAC and members of the LWS consortium in the UK for
their advice on LWS data-reduction techniques. We also thank members of the helpdesk at
Vilspa for their support. Michael Werner, the paper's referee made a number of clearly helpful
comments. M. H. is pleased to acknowledge NASA grant NAG5-3347 to Cornell University;
D. A. N has been supported by NASA grant NAG5-3316. G. J. M. would like to cite support
through NASA LTSA grant NAG5-3542 and contract NAS5-30702, and M. J. K. is pleased to
acknowledge support from NASA RTOP 344-04-10-02 and RTOP 864-03-08-90. \vfill\eject
\centerline{\bf References} \vskip 0.1 true in  {\hoffset 20pt \parindent = -20pt
  
Beck, S. C. 1984, ApJ, 281, 205
  
Beckwith, S., Persson, S. E., Neugebauer, G., \& Becklin, E. E. 1978, ApJ, 223, 464
  
Cardelli, J. A., Meyer, D. M., Jura, M., \& Savage, B. D. 1996, ApJ, 467, 334
  
Cernicharo, J. Gonz\'{a}les-Alfonso, E., Alcolea, J. Bachiller, R., \& John D. 1994, ApJ, 432,
L59
  
Chernoff, D. F., Hollenbach, D. J., \& McKee, C. F. 1982, ApJ, 259, L97
  
Clegg, P. E. {\it et al.} 1996, A\&A 315, L36
 
Draine, B. T., 190, ApJ 241, 1021
  
Draine, B. T., \& Roberge, W. G. 1982, ApJ, 259, L91
 
Elitzur, M., \& de Jong, T. 1973 A\&A, 67, 323
  
Elitzur, M., \& Watson, W. D. 1978, A\&A, 70, 443
  
Kaufman, M. J. 1995, PhD Thesis, Johns Hopkins University
  
Kaufman, M. J., \& Neufeld, D. A. 1996, ApJ, 456, 611
  
Kessler, M. F., Steinz, J. A., Anderegg, M. E., Clavel, J., Drechsel, G., Estaria, P., Faelker, J.,
Riedinger, J. R., Robson, A., Taylor, B. G., \& Xim'{e}nez de Ferr\'{a}n, S. 1996, A\&A,
315, L27
  
Kwan, J., \& Scoville, N. Z. 1977, ApJ, 210, L39
  
Melnick, G. J., Stacey, G. J., Genzel, R., Lugten, J. B., \& Poglitch, A. 1990, ApJ, 348, 161
  
Neufeld, D. A., \& Kaufman, M. J. 1993, ApJ, 418, 263
  
Neufeld, D. A., \& Melnick, G. J. 1987, ApJ, 332, 266
  
Pollack, J. B., Hollenbach, D., Beckwith, S., Simonelli, D. P., Roush, T., \& Fong, W. 1994,
ApJ, 421, 615
  
Stacey, G. J., Kurtz, N. T., Smyers, S. D., \& Harwit, M. 1983, MNRAS, 202, 25p
  
Sugai, H., Usuda, T., Kataza, H., Tanaka, M., Inoue, M. Y., Kawabata, H., Takami, H., Aoki,
T., \& Hiromoto, N. 1994, ApJ, 420, 746
  
Viscuso, P. J., Stacey, G. J., Fuller, C. E., Kurtz, N. T., \& Harwit, M. 1985, ApJ, 296, 142
  
Watson, D. M., Storey, J. W. V., Townes, C. H., Haller., E. E., \& Hansen, W. L. 1980, ApJ,
239, L129
  
Watson, D. M. 1982, PhD Thesis, University of California, Berkeley.

Watson, D. M., Genzel, R., Townes, C. H., \& Storey, J. W. V. 1985, ApJ, 298, 316
  
Werner, M. W., Gatley, I., Harper, D. A., Becklin, E. E., Loewenstein, R. F., Telesco, C. M.,
\& Thronson, H. A., 1976, ApJ, 204, 420 
  
Zmuidzinas, J., Blake, G.A., Carlstrom, J., Keene, J., Miller, D., Schilke, P., and Ugras, N.G.
1995, in {\it Proceedings of the Airborne Astronomy Symposium on the Galactic Ecosystem:  From Gas to Stars to Dust}, eds. M. R. Haas, 
J. A. Davidson \& E. F. Erickson (San
Francisco: ASP), p. 33}

{\scriptsize \vfill\eject \begin{tabular}{l l r c c c c l c} \multicolumn{9}{c}{Table 1}\\
\multicolumn{9}{c}{LWS lines observed toward Orion BN/KL}\\ \hline \hline Line &&
$\lambda$ & $V_{\rm LSR}$ & $\Delta V$ & Line Flux & Line Flux/ & Inferred &
Continuum \\ &&&&&&KN-CMJS& Abundance & Flux \\ && ($\mu$m) & (km s$^{-1}$) &
(km s$^{-1})$ & (W cm$^{-2}$)& Prediction & n(H$_2$O)/n(H$_2$) & (Jy)\\ \hline
  
$4_{04}- 3_{13}$ & Para & 125.3537 & 21 & 70 & $2.2 \times 10^{-17}$ & 1.0 & $3.5
\times 10^{-4}$ & $6.2 \times 10^4$ \\ $4_{32}- 4_{23}$ & Ortho& 121.7219 & 27 & 65 &
$2.2 \times 10^{-17}$ & 5.2& $1.8 \times 10^{-3}\,\,\,\,^a$ & $8.5 \times 10^4$ \\ $5_{05}-
4_{14}$ & Ortho& 99.4931 & 17 & 70 & $7.4 \times 10^{-17}$ & 1.4 & $4.9 \times
10^{-4}$ & $1.5 \times 10^5$ \\ $5_{15}- 4_{04}$ & Para & 95.6273 & 23 & 50 & $2.8
\times 10^{-17}$ & 1.5 & $5.3 \times 10^{-4}$ & $1.6 \times 10^5$ \\ $3_{22}- 2_{11}$ &
Para & 89.9883 & 33 & 54 & $3.7 \times 10^{-17}$ & 1.85 & $6.5 \times 10^{-4}$ & $1.6
\times 10^5$ \\ $6_{06}- 5_{15}$ & Para & 83.2840 & 27 & 45 & $1.6 \times 10^{-17}$ &
1.3 & $4.6 \times 10^{-4}$ & $2.0 \times 10^5$ \\ $6_{16}- 5_{05}$ & Ortho & 82.0315 &
22 & 61 & $4.8 \times 10^{-17}$ & 1.2 & $4.2 \times 10^{-4}$ & $2.2 \times 10^5$ \\
$7_{07}- 6_{16}$ & Ortho & 71.9470 & 23 & 64 & $2.6 \times 10^{-17}$ & 1.1 & $3.8
\times 10^{-4}$ & $2.0 \times 10^5$ \\
  
\hline  \\ \end{tabular}
  
{\noindent $^a$ This abundance estimate should be regarded as an upper limit, because the
121.7~$\mu$m line flux is significantly enhanced by radiative pumping, a process neglected in
the KN model.}}

\vfill\eject \centerline{\bf Figure Caption} \vskip 0.1 true in  {\hoffset -20pt \parindent -20pt
Fig. 1 -- Line profiles of the observed transitions listed in Table 1. Error bars showing standard
deviations from a mean obtained in successive spectral scans are too small to be discerned on
this scale. Corrections, however, were made by eliminating grossly deviant data points
resulting from cosmic ray impacts on detectors. Given the large number of scans obtained,
these subtractions did not significantly alter the final results.}

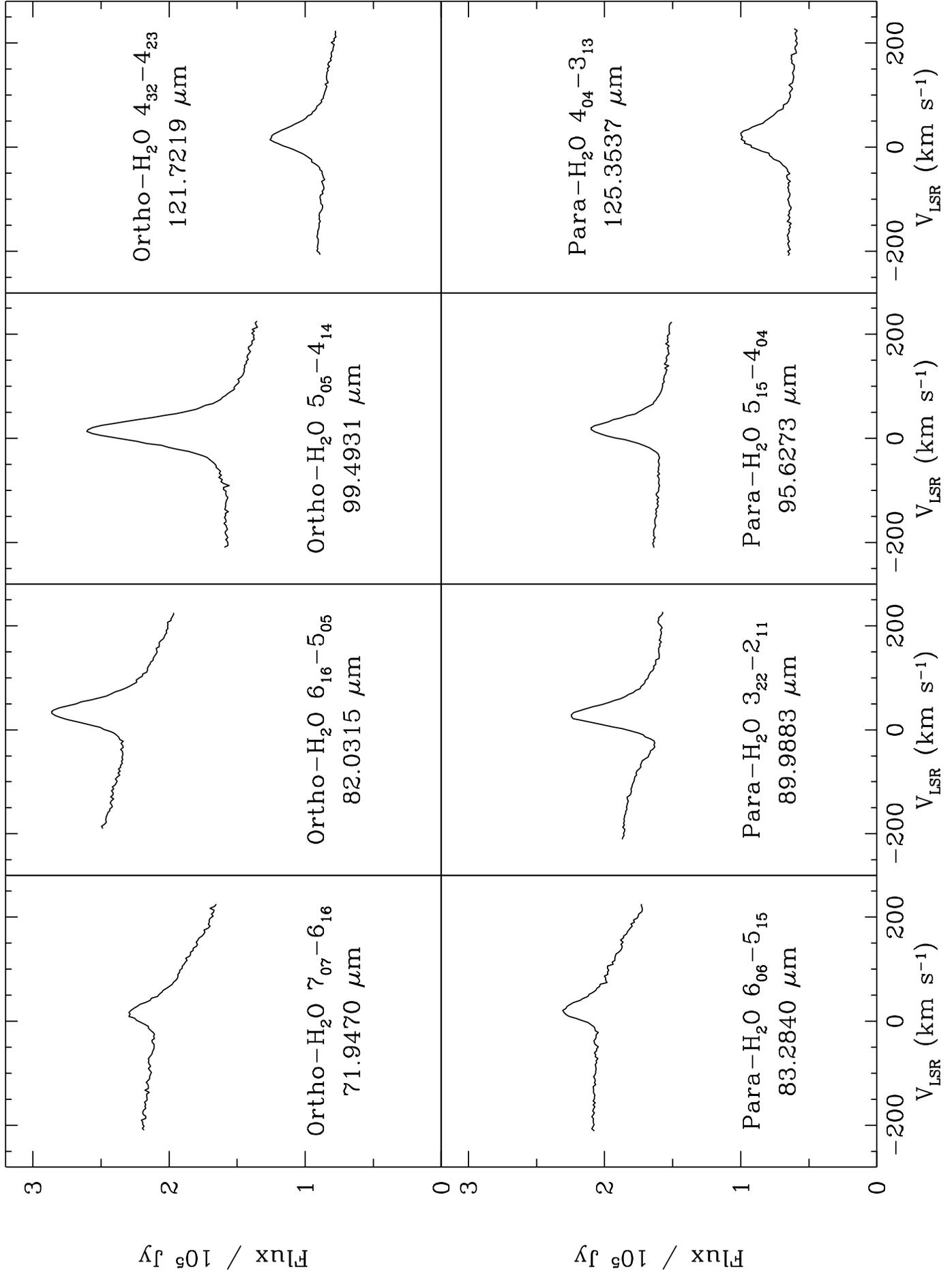
\end{document}